\documentclass[aps,twocolumn]{revtex4-1}
\usepackage{graphicx}
\usepackage{latexsym}
\usepackage{wasysym}
\usepackage{amssymb}

\begin{document}

\pacs{87.15.A-, 36.20.Ey, 87.15.H-}
\title{Accuracy of the blob model for single flexible polymers inside nano-slits that are a few monomer sizes wide}

\author{Narges Nikoofard}
\email{nikoofard@kashanu.ac.ir}
\affiliation{Institute of Nanoscience and Nanotechnology, University of Kashan, Kashan 51167-87317, Iran}

\author{S. Mohammad Hoseinpoor}
\affiliation{Institute of Nanoscience and Nanotechnology, University of Kashan, Kashan 51167-87317, Iran}

\author{Mostafa Zahedifar}
\affiliation{Institute of Nanoscience and Nanotechnology, University of Kashan, Kashan 51167-87317, Iran}

\date{\today}

\begin{abstract}
The de Gennes' blob model is extensively used in different problems of polymer physics. This model is theoretically applicable when the number of monomers inside each blob is large enough. For confined flexible polymers, this requires the confining geometry to be much larger than the monomer size. In this manuscript, the opposite limit of polymer in nano-slits with one to several monomers width is studied, using molecular dynamics simulations. Extension of the polymer inside nano-slits, confinement force on the plates, and the effective spring constant of the confined polymer are investigated. Despite of the theoretical limitations of the blob model, the simulation results are explained with the blob model very well. The agreement is observed for the static properties and the dynamic spring constant of the polymer. A theoretical description of the conditions under which the dynamic spring constant of the polymer is independent of the small number of monomers inside blobs is given. Our results on the limit of applicability of the blob model can be useful in the design of nano-technology devices.

\end{abstract}

\maketitle

\section{Introduction}

Advances in nanotechnology have enabled investigations of single polymers confined in the nano-scale \cite{nature2006,advances}. This problem has attracted interests due to its applications in the design of nanotechnology devices \cite{translocation,separation,barcode}. It also helps in obtaining more knowledge of polymers confined in biological environments \cite{genome,jun_pnas2006}. Besides, it has become possible to check accuracy of the old theories and improve them \cite{austin,slit-systematic}. Single polymers confined inside nano-spheres \cite{ejtehadi,gao2014}, nano-channels \cite{austin,arnold,jung,soft2013,advances} and nano-slits (between two parallel plates) \cite{dekker,binder2008,slit-systematic,slit-revisit,semiflexible} are studied. 

The most known theory for confined polymers inside nano-channels or nano-slits is the de Gennes' blob model. For semi-flexible polymers, Odijk theory is relevant when the size of the confining geometry, $D$, is smaller than the persistence length of the polymer, $L_p$ \cite{austin,dekker}. Recently, it is shown that extended de Gennes' theory should be used in the case of semi-flexible polymers when $L_p < D < \frac{L_p^2}{w}$ \cite{extended}. Here, $w$ is the monomer width.

Most of the studies on confined polymers focus on double-stranded DNA, which is a semi-flexible polymer. Recently, single stranded DNA is introduced as a model system for study on flexible polymers \cite{ssDNA}. Also, many proteins and synthetic polymers are flexible.
For flexible polymers, the only relevant theory is the de Gennes' blob model and it would be useful to better understand its limitations and regime of applicability. 
The blob model has broad applications in different problems of polymer physics, such as polyelectrolytes \cite{joanny}, polymers under tension \cite{tension}, and polymers adsorbed to surfaces \cite{pre2013} or confined near surfaces unfer external fields \cite{nikoofard}. 

In the blob model, the confined polymer is divided into smaller sections, called $blobs$, that have the size of the confining geometry, $D$. Inside blobs, polymer does not feel that it is confined. The number of monomers inside each blob is found from the relations of the self-avoiding walk (SAW) in free space, $g\sim \left(\frac{D}{b}\right)^{\frac{1}{\nu}}$. Here, $\nu$ is the Flory exponent in 3D and $b$ is the monomer size. In length scales larger than the blob size, the polymer (with $N$ real monomers) behaves like an effective polymer with $\frac{N}{g}$ monomers of size $D$. This effective polymer has a SAW in one or two dimensions for a nano-channel or a nano-slit, respectively. The number of monomers inside each blob, $g$, and total number of the blobs, $\frac{N}{g}$, should be large enough, to be able to use the relations of statistical mechanics for SAW inside or outside the blobs, respectively \cite{rubinstein}.

A number of studies on confined polymers are performed in conditions that the above two requirements are not satisfied. It is shown that the blob model is applicable in micro-devices, where the number of blobs is very small \cite{pre2010}. Also in some practical situations, the polymer is confined in severe conditions that the number of monomers inside each blob is very small \cite{severe}. Specially in nano-devices for DNA barcoding, full stretching of DNA is required \cite{barcode}. Computer simulations show that the blob model can explain the statics of a flexible polymer inside narrow channels  \cite{arnold,soft2013}. However, the polymer dynamics in nano-channels is only described with the blob model when the number of monomers inside each blob is large enough \cite{jung,soft2013}. 

In this manuscript, a flexible polymer confined inside a nano-slit is studied using molecular dynamics (MD) simulations. The extreme condition that the number of monomers inside each blob is small is considered. It is shown that our results for the radius of gyration and confinement free energy of the polymer can be explained very well, using the blob model. Despite previous studies (Ref. \cite{binder2008} and references therein), no correction is made to the slit width (distance between the two plates). Agreement between simulation results and the blob model is observed even for very narrow slits with 1-2 monomer(s) width. These widths are completely below the regime of applicability of the blob model.

Extension of the polymer inside the nano-slit has fluctuations around its mean value. From these fluctuations, an effective spring constant can be obtained. It is a dynamic variable for the polymer and is a measure of elasticity of the polymer. Our results for the effective spring constant are in good agreement with the blob model, unlike the case of the polymer inside the nano-channel \cite{arnold,jung,soft2013}. 

This discrepancy is due to the difference between the origins of the effective spring constant of a polymer confined in a nano-channel and a nano-slit. Two factors govern the fluctuations of a polymer confined in a nano-slit: fluctuations in the size of blobs  and fluctuations in the self-avoiding walk of the blobs. It is obvious that the latter factor is dominant. However, for a polymer confined in a nano-channel, the blobs are arranged one after the other along the channel. So, fluctuations in the blob size are the only source of fluctuations in the extension of the polymer (see Fig. \ref{fig0}). This makes the internal dynamics of the blobs important, in the dynamics of the whole polymer. As a result, it is determining to have the sufficient number of monomers inside each blob, for a polymer confined inside a nano-channel.

In the next section, the statics and dynamics of a flexible polymer confined in a nano-slit is reviewed briefly. In Sec \ref{method}, our simulation method is described. Results of the simulation and their compatibility with the theory are discussed in Sec. \ref{results}. Finally, a summary of the results is presented in the last section.

\section{Theory}    \label{theory}

\begin{figure}
\includegraphics[scale=0.2]{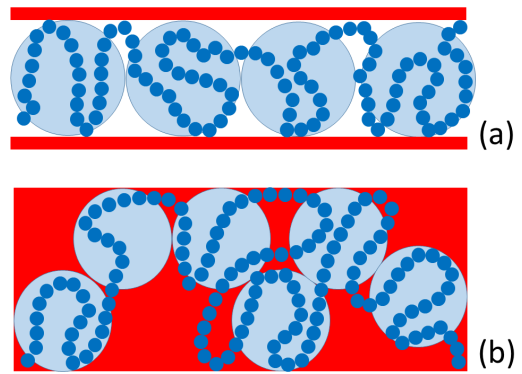}
\caption{(Color online) (a) Blob model for a flexible polymer confined inside a nano-channel. The blobs of the polymer in the nano-channel are arranged one after the other along the nano-channel. (b) Top view of a polymer confined inside a nano-slit. The blobs of the polymer in the nano-slit span the 2D space between the plates. The number of monomers inside one blob and total number of the blobs are clear in the two cases (a) and (b).  }
\label{fig0}
\end{figure}

Consider a flexible polymer confined between two parallel plates  (Fig. \ref{fig0}). Confinement extends the polymer on the walls, to a radius of gyration larger than its free radius of gyration $R_g\sim bN^{\nu}$. Here, $N$ is the number of monomers of the polymer, $b$ is the monomer size and $\nu=\frac{3}{5}$ is the Flory exponent in 3D. Sections of the chain smaller than the distance between the plates, $D$, do not feel the constraints. These sections, called blobs, have a statistics similar to a chain in free space $D \sim bg^{\frac{3}{5}}$, where $g$ is the number of monomers inside each blob.

The polymer is composed of $\frac{N}{g}$ blobs that cannot penetrate into each other. So, it can be regarded as an effective two-dimensional polymer with $\frac{N}{g}$ monomers of size $D$. We use the exponent of a self-avoiding walk in two dimensions, $\nu=\frac{3}{4}$, to find the extension of the polymer parallel to the plates $R_{||}\sim D(\frac{N}{g})^{\frac{3}{4}}$. This gives the relation between the average extension of the polymer with the total number of monomers and the distance between the plates,

\begin{equation} \label{equi-ext}
R_{||}\sim b N^{\frac{3}{4}} \left(\frac{D}{b}\right)^{-\frac{1}{4}}.
\end{equation} 

Free energy of confining the polymer is $k_BT$ per blob, $F_{conf} \approx k_BT\frac{N}{g}$. Substituting $g$ in this relation gives 

\begin{equation} \label{equi-energy}
F_{conf} \approx k_BTN\left(\frac{b}{D}\right)^{\frac{5}{3}}.
\end{equation}

The force exerted on the plates is found by differentiating the confinement free energy with respect to D,
\begin{equation} \label{force}
f\approx \frac{k_BT}{b} N \left(\frac{b}{D}\right)^{\frac{8}{3}}.
\end{equation} 

The polymer has fluctuations around its average extension. The mean square deviation from the average extension is described by an effective spring constant $\langle\Delta R_{||}^2\rangle=\langle\left(R_{||}(t)-R_{||av}\right)^2\rangle\approx k_BT/k_{eff}$. The simplest way to calculate the spring constant, $k_{eff}$, is to eliminate $D$ between the relations \ref{equi-ext} and \ref{equi-energy}. This gives the free energy as a function of extension. Differentiating the polymer free energy twice with respect to its extension and substituting the equilibrium extension gives a spring constant \cite{austin}. This calculation does not give a correct result, because equation \ref{equi-energy} is the free energy of the polymer at its equilibrium extension. Indeed, the free energy of the polymer for any desired extension should be used in the calculation. 

Suppose the two ends of the confined polymer are under tension and it is elongated to the extension $R_f$. Free energy of a two- or three-dimensional polymer with $N$ monomers of size $b$ which is extended to size $R_f$ is $F_{tens} \approx k_BT\left(\frac{R_f}{bN^{\nu}}\right)^{\frac{1}{1-\nu}}$ \cite{rubinstein}. $\nu$ is equal to $\frac{3}{5}$ for 3D and equal to $\frac{3}{4}$ for 2D. By differentiating this free energy twice with respect to $R_f$ and substituting the equilibrium extension $R_{eq}\sim bN^{\nu}$, the spring constant is obtained as $k_{eff}=\frac{\partial^2F_{tens}}{\partial R_f^2}|_{R_{eq}}\approx\frac{k_BT}{(bN^{\nu})^2}\approx \frac{k_BT}{R_{eq}^2}$. This is in agreement with the relation used in Ref. \cite{spring}.

Here, the confined polymer is described as an effective polymer with $\frac{N}{g}$ monomers of size $D$ in two dimensions ($\nu=\frac{3}{4}$). So, the spring constant becomes $k_{eff}\approx \frac{k_BT}{\left(D\left(\frac{N}{g}\right)^{\frac{3}{4}}\right)^2}$. Substituting for $g$, we get the spring constant for the confined polymer 
\begin{equation} \label{spring}
k_{eff}\approx \frac{k_BT}{b^2}N^{-3/2}\left(\frac{D}{b}\right)^{\frac{1}{2}}=\frac{k_BT}{R_{||}^2}.
\end{equation}
In the final relation, $R_{||}$ is substituted from eq. \ref{equi-ext}. To derive this relation, the confined polymer is simply modeled as  an effective two-dimensional polymer composed of blobs and the effect of the external tension is to move the confinement blobs toward a linear alignment. This simple description is valid for small tensions with $R_f < D(\frac{N}{g})$.

The blob model is accurate when the number of monomers inside each blob, $g$, and total number of the blobs, $\frac{N}{g}$, are considerable. This is necessary to be able to use statistical mechanics relations for a self-avoiding walk inside and outside the blobs, respectively. The size spanned by a self-avoiding walker in space is proportional to the number of steps to the power of the Flory exponent in two or three dimensions, when the number of steps is large enough. For $g$ to be large enough, the confining geometry (nano-slit or nano-channel) should be much larger than one monomer size. 

It is instructive to compare the effect of an external tension on a polymer confined inside a nano-channel or a nano-slit. Prior to tension, the polymer inside the nano-slit arranges its blobs in two dimensions. The effect of the external tension is to move the blobs toward a one dimensional arrangement. However, the blobs of the polymer inside a nano-channel  have already lost two degrees of freedom and lie on a one dimensional line. So, the polymer has to stretch its blobs to respond to the external tension. One can conclude that the elasticity of a polymer inside a nano-slit has its origin in the arrangement of the blobs. This is in contrast to the elasticity of a polymer inside a nano-channel which originates from the elasticity of each blob.
As a result, to use the blob model for the former case, the number of blobs should be sufficient; but for the latter case, the number of monomers inside each blob is also important.

\section{Simulation Method} \label{method}

\begin{figure}
\includegraphics[scale=0.2]{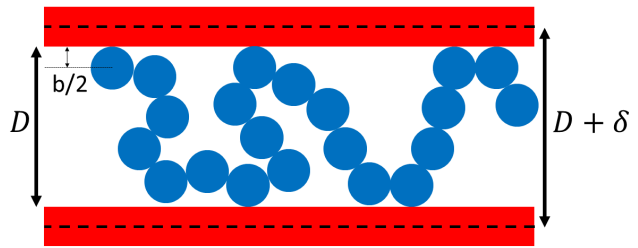}
\caption{(Color online) Side view of a polymer confined in a nano-slit. The surface of each plate is defined such that the closest distance of the center of a monomer to the plate is equal to $b/2$. Distance between the plates in our paper, $D$,  is shown in the figure. َAn effective width for the nano-slit is also shown, to examine the effect of adding correction to the slit width.}
\label{fig1}
\end{figure}

We use MD simulations to check the limits of accuracy of the blob model for a flexible polymer confined in a nano-slit. A schematic of the system is shown in Fig. \ref{fig1}. The polymer is modeled as a bead-spring chain. Monomers of the polymer are connected by the FENE potential,
\begin{equation}\label{fene}
U_{FENE}(r)= -\frac{1}{2} K r_0^2 \ln\left[1-\left(\frac{r}{r_0}\right)^2\right].
\end{equation}
Shifted-truncated Lennard-Jones is used for the excluded volume interactions between the monomers and between the monomers and the plates,
\begin{equation}\label{LJ}
U_{LJ}(r)= 4\epsilon\left[\left(\frac{b}{r}\right)^{12}-\left(\frac{b}{r}\right)^{6}+\frac{1}{4}\right],  \qquad r< 2^{\frac{1}{6}}b.
\end{equation}
$\epsilon$ and $b$ are the energy and length scales of the simulation. $K=100\epsilon$ and $r_0=1.5b$ are used for the FENE potential.

The plates are infinite and the closest distance of the center of a monomer to the plate surface is $b/2$. This means that the spherical surface of the monomer can touch the plate, in agreement with our intuition from a wall (Fig. \ref{fig1}). In previous studies, the center of a monomers was allowed to touch the impenetrable plate surface \cite{binder2008}. So, it was needed to add the monomer size $b$ to the distance between the walls to compare the results with the blob model. It is important to note that this is not attributed to the finite size of the chains in these simulations. Here, it will be shown that the simulation results for similar chains are in agreement with the blob model, without any corrections in the wall distance.

The equations of motion are integrated using the Velocity Verlet algorithm, with the step size equal to $0.01\tau_0$. $\tau_0=\sqrt{\frac{m\sigma^2}{\epsilon}}$ is the MD time scale and $m$ is the monomer mass. The system is kept at the constant temperature $T=1.0\frac{\epsilon}{k_B}$, using the Langevin thermostat with the friction coefficient $1.0 \tau_0^{-1}$. The simulations are performed using ESPResSo \cite{espresso}.

In our simulations, the nano-slit width is changed from $b$ to 9$b$. Polymers with 100, 200 and 300 monomers are investigated. For slits with width 3-9$b$, the monomers are arranged in an initial structure close to a self-avoiding walk. Then, the system is warmed up to remove any overlap between the monomers. For slits with width $b$ and 2$b$, the monomers are initially arranged on one line. This initial configuration reaches equilibrium in a longer time interval, but it is inevitable because of the very small width of the slits. 

All simulations are performed for a total time of $5N^2\tau_0$. Data points to find time averages of the static and dynamic properties of the polymer are collected after the time $N^2\tau_0$, in time intervals equal to $\tau_0$. In all plots, the size of errorbar at each point is smaller than the symbol size.

The radius of gyration of the polymer parallel to the plates is calculated using the relation $R_{||}=\frac{1}{N}\sqrt{(X_i-X_{CM})^2+(Y_i-Y_{CM})^2}$. $X_i$ and $X_{cm}$ are the x components of the positions of the ith monomer and the center of mass, respectively. $Y_i$ and $Y_{CM}$ are the corresponding y components. The force on each plate in all steps of the simulation is sum of the Lennard-Jones forces from all monomers interacting with the plate. 

\begin{figure}
\includegraphics[scale=0.48]{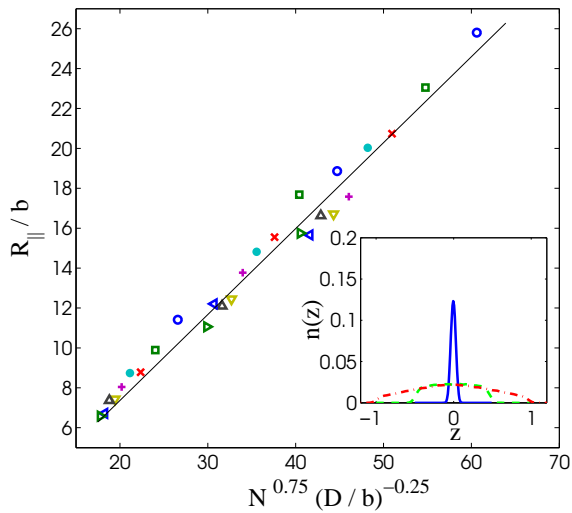}
\caption{(Color online) Simulation results for $R_{||}$ versus the theoretical relation $N^{0.75}D^{-0.25}$. $\Circle$,  $\Box$, $\times$, $\CIRCLE$, $+$, $\triangledown$, $\vartriangle$, $\vartriangleleft$ and $\vartriangleright$ show the simulation results for D/b=1-9, in ascending order. Inset: Distribution of the monomers in the distance between the plates. The solid, dashed and dash-dotted lines correspond to the three narrowest slits D/b=1-3, in ascending order.}
\label{fig2}
\end{figure}

\begin{figure}
\includegraphics[scale=0.9]{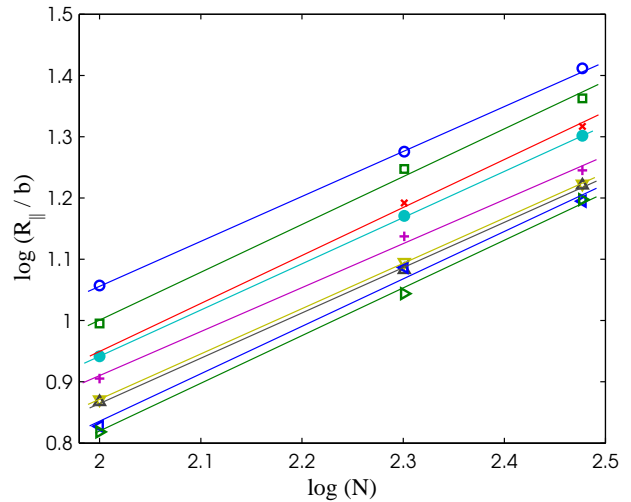}
\caption{(Color online) log$_{\text{10}}$-log$_{\text{10}}$ plot of radius of gyration of the polymer parallel to the plates $R_{||}$ versus the polymer length, $N$, for different nano-slit widths.
$\Circle$,  $\Box$, $\times$, $\CIRCLE$, $+$, $\triangledown$, $\vartriangle$, $\vartriangleleft$ and $\vartriangleright$ show the simulation results for D/b=1-9, in ascending order. The corresponding $\alpha$ values are 0.74, 0.78, 0.79, 0.76, 0.72, 0.74, 0.74, 0.78 and 0.79.
The average of $\alpha$ is 0.76.}
\label{fig3}
\end{figure}

\begin{figure}
\includegraphics[scale=0.9]{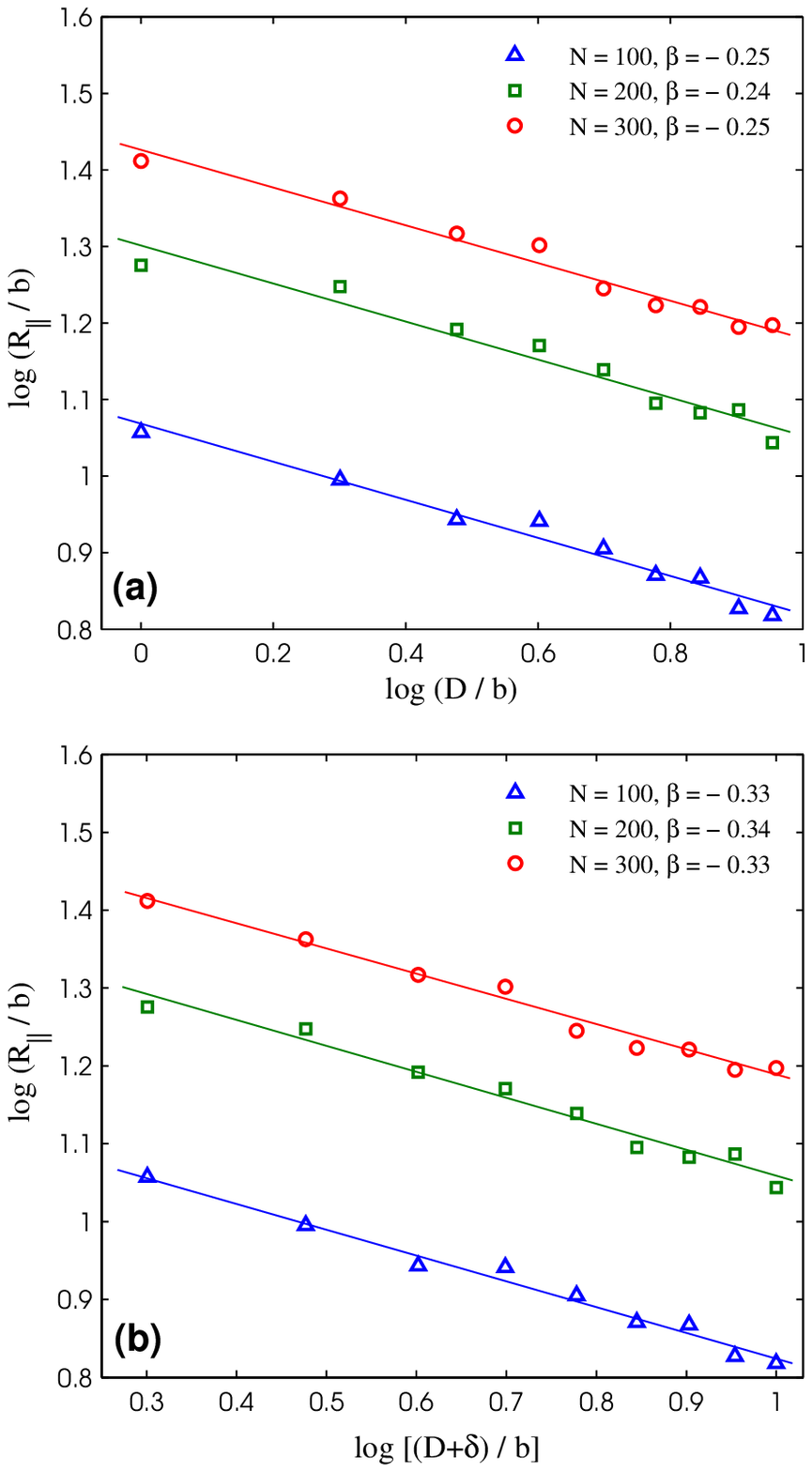}
\caption{(Color online) log$_{\text{10}}$-log$_{\text{10}}$ plot of  $R_{||}$ versus the nano-slit width. (a) and (b) show two different definitions for the nano-slit width (see Fig. \ref{fig1}).
As can be seen, omitting $\delta$ results in an excellent improvement in the scaling exponent.}
\label{fig4}
\end{figure}

\begin{figure}
\includegraphics[scale=0.48]{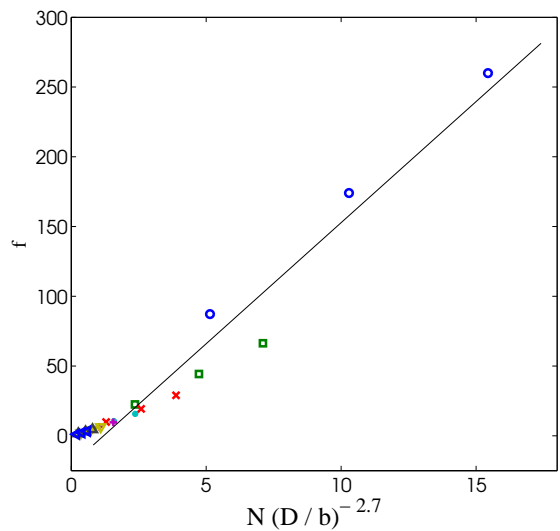}
\caption{(Color online) Simulation results for $f$ versus the theoretical relation $N D^{-2.7}$. $\Circle$,  $\Box$, $\times$, $\CIRCLE$, $+$, $\triangledown$, $\vartriangle$, $\vartriangleleft$ and $\vartriangleright$ show the simulation results for D/b=1-9, in ascending order.}
\label{fig5}
\end{figure}

\begin{figure}
\includegraphics[scale=0.9]{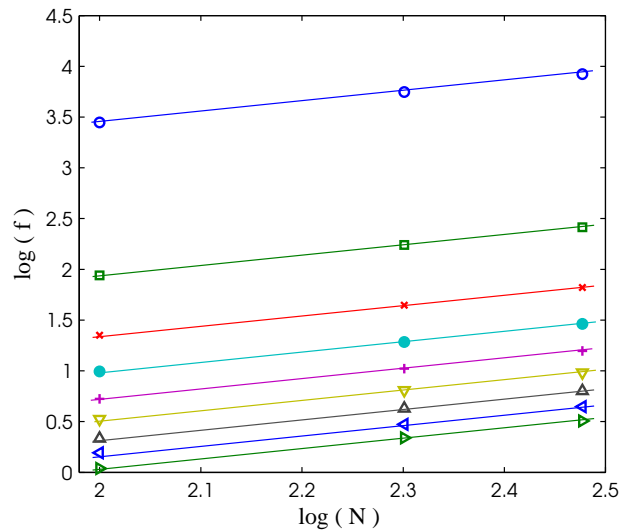}
\caption{(Color online) log$_{\text{10}}$-log$_{\text{10}}$ plot of $f$ versus the polymer length, $N$, for different nano-slit widths. $\Circle$,  $\Box$, $\times$, $\CIRCLE$, $+$, $\triangledown$, $\vartriangle$, $\vartriangleleft$ and $\vartriangleright$ show the simulation results for D/b=1-9, in ascending order. The corresponding $\gamma$ values are 1, 0.99, 0.99, 0.98, 0.99, 0.96, 0.98, 0.95 and 0.99.
 The average of $\gamma$ is 0.99.}
\label{fig6}
\end{figure}

\section{Simulation Results} \label{results}

\subsection{Radius of gyration}

Radius of gyration of the polymer parallel to the plates versus the theoretical relation in equation \ref{equi-ext} is shown in Fig. \ref{fig2}. All data lie on a single line, in the whole range of parameters studied. This shows agreement between the simulation results and the blob model. Distribution of the monomers in the distance between the plates is shown in the inset of Fig. \ref{fig2}, for the three narrowest slits. Because of the soft nature of the potential, monomers can penetrate the walls rarely.

To further investigate the correspondence of the simulation results with the theory, exponents $\alpha$ and $\beta$ in the relation $R_{||}\propto N^{\alpha}D^{\beta}$ are obtained from simulation. These exponents are found from the slopes of the log$_{\text{10}}$-log$_{\text{10}}$ plots of the radius of gyration versus the polymer length and the nano-slit width, respectively.  As can be seen in Fig. \ref{fig3}, $\alpha$ is close to the theoretical value even for very narrow slits (1-2$b$). $\alpha$ is not a decreasing or increasing function of the nano-slit width. So, the small number of monomers inside each blob is not determining in this exponent. The average of $\alpha$ is 0.76, which has \%1 deviation from the theoretical value. Fig. \ref{fig4}(a) shows the exponent $\beta$ for different lengths of the polymer. It is seen that finite size of the chains has no effect on the simulation results. The results are summarized in the relation $R_{||}\propto N^{0.76}D^{-0.25}$.

For comparison with Ref. \cite{binder2008}, an effective width is defined for the nano-slit by adding a monomer size to the nano-slit width (see Fig. \ref{fig1}) in Fig. \ref{fig4}(b). The exponent $\beta$ deviates largely from the theoretical value, with this change in the definition of the wall surface.

\begin{figure}
\includegraphics[scale=0.9]{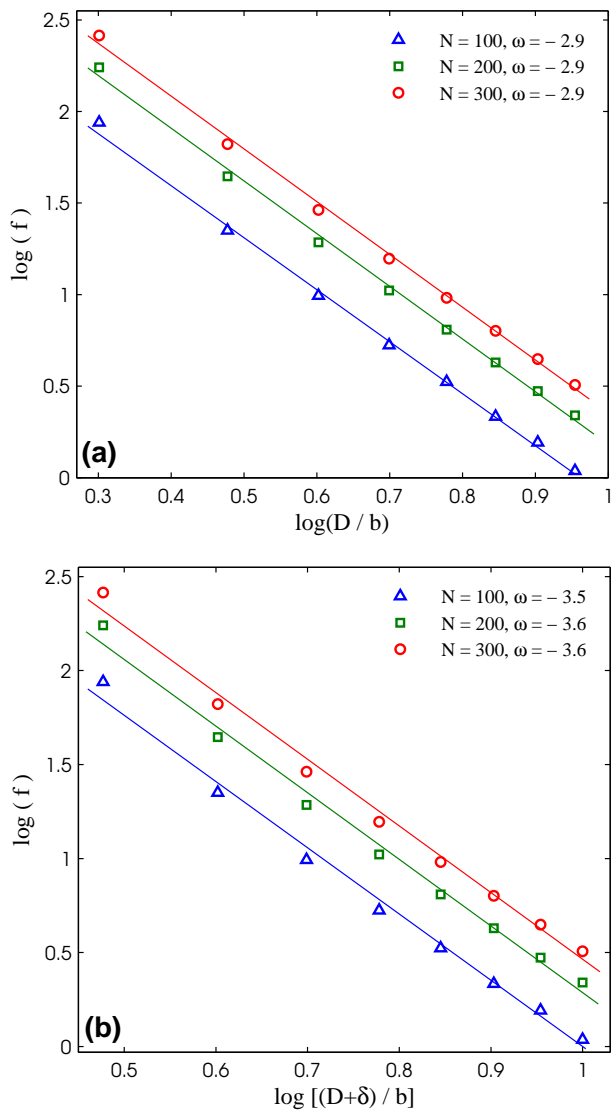}
\caption{(Color online) log$_{\text{10}}$-log$_{\text{10}}$ plot of  $f$ versus the nano-slit width. (a) and (b) show two different definitions for the nano-slit width (see Fig. \ref{fig1}).
 As can be seen, omitting $\delta$ results in an excellent improvement in the scaling exponent. }
\label{fig7}
\end{figure}

\subsection{The confinement force}

It is not easy to measure the free energy of confining the polymer in the nano-slit. Instead, the force that the confined polymer exerts on the plates is studied in the simulation. The results are compared with equation \ref{force} from the theory. The simulation results for the force are plotted versus the theoretical relation in Fig. \ref{fig5}. A linear behavior is observed, which shows agreement between theory and simulation. 

The exponents $\gamma$ and $\omega$ in the relation $f\propto N^{\gamma}D^{\omega}$ are obtained from simulation.  According to Fig. \ref{fig6}, deviation in $\gamma$ from the theoretical value (=1) is less than \%1. 

In Fig. \ref{fig7}(a), the value of $\omega$ is found from the log$_{\text{10}}$-log$_{\text{10}}$ plot of force versus the nano-slit width. Deviation of the exponent from the theoretical value (-2.7) is less than \%8. The simulation results are summarized in $f\propto N^{0.99}D^{-2.9}$. Again, to check our definition for the nano-slit width, the exponent $\omega$ is found from the log$_{\text{10}}$-log$_{\text{10}}$ plot of force versus $D+\delta$ (Fig. \ref{fig7}(b)). Deviation of the exponent becomes more than \%30.

\subsection{The effective spring constant}

\begin{figure}
\includegraphics[scale=0.48]{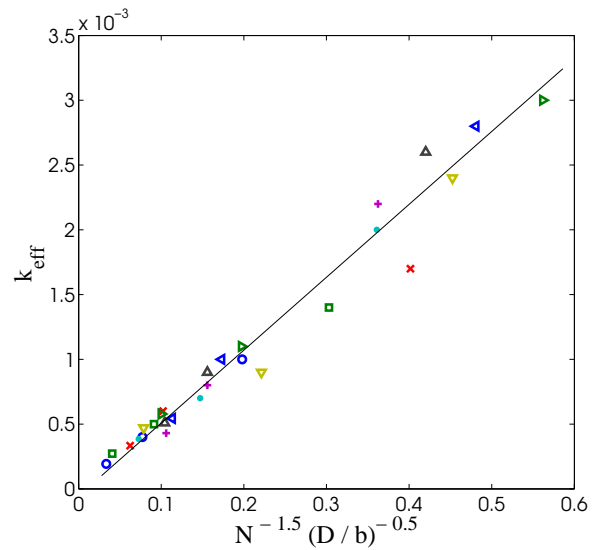}
\caption{(Color online) Simulation results for the effective spring constant versus the theoretical relation $N^{-1.5}D^{0.5}$. $\Circle$,  $\Box$, $\times$, $\CIRCLE$, $+$, $\triangledown$, $\vartriangle$, $\vartriangleleft$ and $\vartriangleright$ show the simulation results for D/b=1-9, in ascending order.}
\label{fig8}
\end{figure}

\begin{figure}
\includegraphics[scale=0.9]{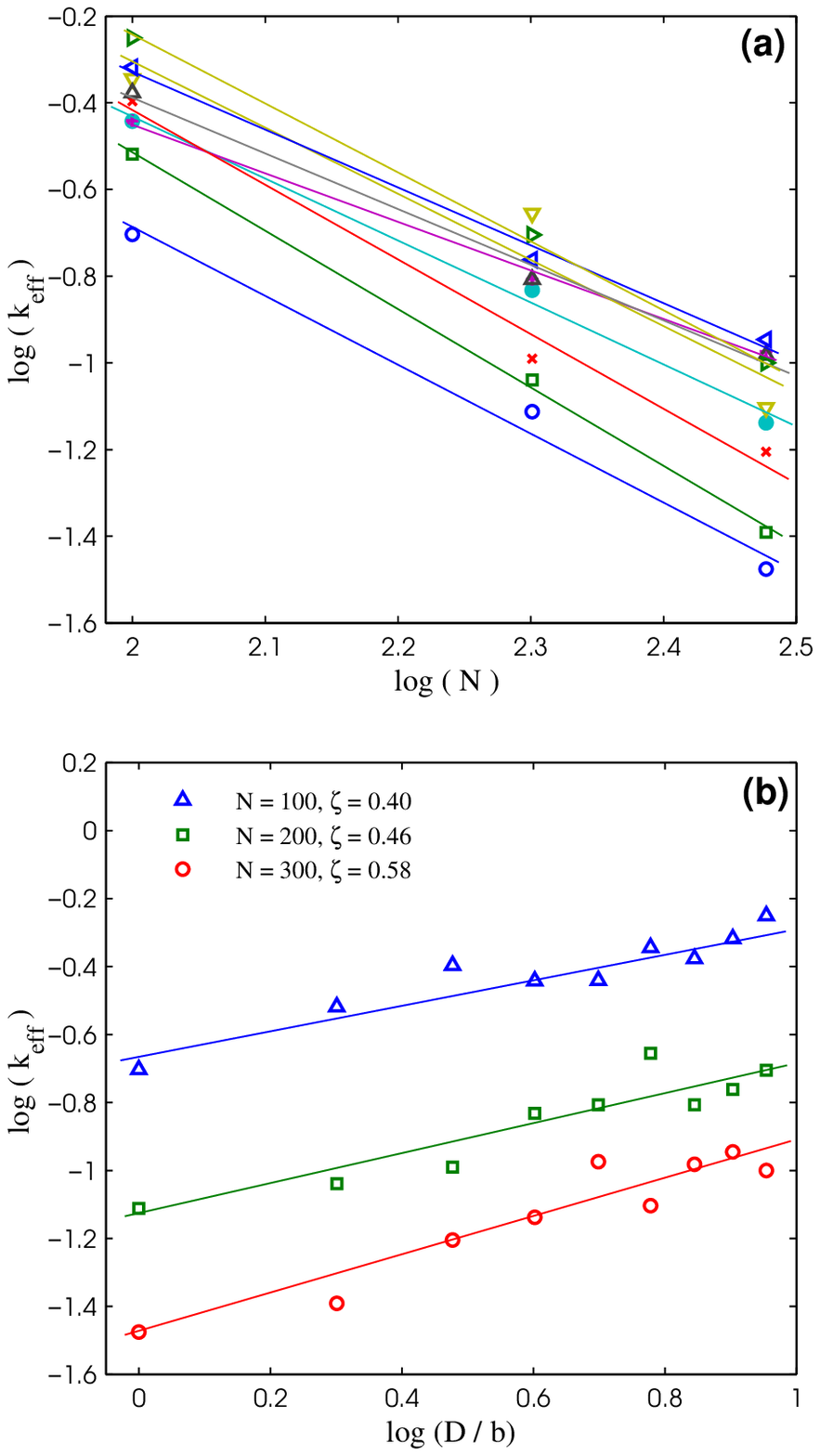} 
\caption{(Color online) (a) log$_{\text{10}}$-log$_{\text{10}}$ plot of $k_{eff}$ versus the polymer length, $N$, for different nanoslit width. $\Circle$,  $\Box$, $\times$, $\CIRCLE$, $+$, $\triangledown$, $\vartriangle$, $\vartriangleleft$ and $\vartriangleright$ show the simulation results for D/b=1-9, in ascending order. The corresponding $\eta$ values are -1.6, -1.6, -1.7, -1.4, -1.1, -1.5, -1.3, -1.3, -1.6.
 The average $\eta$ is -1.3. (b) log$_{\text{10}}$-log$_{\text{10}}$ plot of $k_{eff}$ versus the slit width, for different polymer lengths. The average of $\zeta$ is – 0.48. }
 \label{fig9}
\end{figure}

Fluctuations in the radius of gyration $\langle\Delta R_{||}^2\rangle$ can be used to find an effective spring constant for the polymer, $k_{eff}\approx \frac{k_BT}{\langle\Delta R_{||}^2\rangle}$. This effective spring constant is a measure of the dynamics of the confined polymer. For example, the relaxation time of the confined polymer is found by dividing the friction coefficient from the solvent to this effective spring constant \cite{dekker}.

 The simulation results for the effective spring constant versus the theoretical relation (equation \ref{spring}) are depicted in Fig. \ref{fig8}. The linear curve confirms the excellent agreement between theory and simulation. This agreement is observed even when the nano-slit width is very small. 

The exponents $\eta$ and $\zeta$ in the relation $k_{eff} \propto N^{\eta}D^{\zeta}$ are found from the simulation results. Fig. \ref{fig9}(a) shows the log$_{\text{10}}$-log$_{\text{10}}$ plot of $k_{eff}$ versus the polymer length. The average slope of the plots is -1.3. This exponent has the largest deviation, \%13, among the exponents investigated in this manuscript. The exponent is not a monotonic function of the nano-slit width and it cannot be deduced that an improved agreement between theory and simulation would be observed with increasing the nano-slit width. 

Dependence of the effective spring constant on the slit width is investigated in Fig. \ref{fig9}(b). The average of the exponent is 0.48. Although deviation of the average from the theory is around \%4,  the exponents themselves are not very close to it. Overall, a good agreement between theory and simulation is observed and the simulation results are summarized in the relation $k_{eff}\propto N^{-1.3}D^{0.48}$. 

To check the effect of correlations in the calculation of $\langle\Delta R_{||}^2\rangle$, a different averaging scheme was also examined. Data points with a time interval of $100\tau_0$ were collected from the whole data. Starting from different initial times, 100 different series of data points can be defined. The values of the spring constant obtained from each of these series are very close to the ones shown on Figs. \ref{fig9}(a) and \ref{fig9}(b). 

Further investigation is needed to explain the larger deviations of the dynamic exponents from theory, relative to the static ones. One can attribute these deviations to the finite length of the chains or insufficient simulation times for equilibration of the dynamic quantities. However, larger fluctuations may be an intrinsic property of the dynamic quantities and persist even in simulations with longer chains and longer run times.

\section{Summary and Discussion} \label{summary}

In this manuscript, we checked the accuracy of the de Gennes' blob model for a flexible polymer inside a nano-slit, using MD simulation. Theoretically, the blob model is accurate for wide slits and long polymers. However, we examined the model in the limit of very narrow slits. Nano-slits with widths equal to 1-9 times a monomer size were investigated. Simulation results both for the static properties and the dynamic spring constant of the polymer were in excellent agreement with the theory.
 
Our results showed that the blob model can describe the effective spring constant of the confined polymer even for very narrow slits. This is in contrast to a recent study on single flexible polymers inside nano-channels. Indeed, the blob model is only applicable for the spring constant of the polymer in nano-channels with at least 10 monomers width. Here, it was explained that this discrepancy is a result of the different origins of the polymer dynamics in the two cases. The internal dynamics of a blob is determining in the overall dynamics of the chain, for a polymer confined in a nano-channel. But, for a polymer confined in a nano-slit, the dynamics is dominated by the two-dimensional arrangement of the blobs between the plates.

In our simulations, polymers with 100-300 monomers were used.  It was observed that the effect of the finite size of the chains is not considerable. Our results on the accuracy of the blob model even for narrow slits and short chains can be very useful in the studies of polymers in confinement. Polymer confinement occurs in nano-technology devices, such as nano-pore sequencing \cite{translocation,khalilian}, DNA barcoding \cite{barcode} and polymer separation devices \cite{separation}. 

The main result of this manuscript was that the blob model is sometimes accurate beyond its theoretical limits, in practical situations. Considering that the blob model is used in many problems of polymer physics \cite{rubinstein}, this result is applicable in different circumstances. A polymer adsorbed to a surface \cite{pre2013} or compressed on a surface by an external field \cite{nikoofard} is confined to a region near the surface. For these problems with 2D confinement of the polymer, the above result can be used directly. However, for a polymer under external tension more caution is needed, since the blobs are arranged in 1D and the dynamics may be more sensitive to the number of monomers inside one blob.

\begin{acknowledgments}
The authors would like to thank Dr. Hossein Fazli for useful discussions.
\end{acknowledgments}


\begin{thebibliography}{100}

\bibitem{nature2006}
H. Craighead, Nature (London) \textbf{442}, 387 (2006).

\bibitem{advances}
W. Reisner, J. N. Pedersen, and R. H. Austin, Rep. Prog. Phys. \textbf{75}, 106601 (2012).

\bibitem{translocation}
D. P. Hoogerheide, B. Lu, and J. A. Golovchenko, ACS Nano \textbf{8}, 7384 (2014).

\bibitem{separation}
J. Fu, R. B. Schoch, A. L. Stevens, S. R. Tannenbaum, and J. Han, Nature Nanotech. \textbf{2}, 121 (2007).

\bibitem{barcode}
K. Jo \emph{et al}, Proc. Natl. Acad. Sci. USA \textbf{104}, 2673 (2007);
R. Marie \emph{et al}, Proc. Natl. Acad. Sci. USA \textbf{110}, 4893 (2013).

\bibitem{genome}
J. Pelletier \emph{et al}, Proc. Natl. Acad. Sci. USA \textbf{14}, E2649 (2012).

\bibitem{jun_pnas2006}
S. Jun and B. Mulder, Proc. Natl. Acad. Sci. USA \textbf{103}, 12388 (2006).

\bibitem{austin}
W. Reisner \emph{et al}, Phys. Rev. Lett. \textbf{94}, 196101 (2005).

\bibitem{slit-systematic}
L. Dai, J. J. Jones, J. R. C. van der Maarel and P. S. Doyle, Soft Matter \textbf{8}, 2972 (2012).

\bibitem{ejtehadi}
A. Fathizadeh, M. Heidari, B. Eslami-Mossallam, M. R. Ejtehadi, J. Chem. Phys. \textbf{139}, 044912 (2013).

\bibitem{gao2014}
J. Gao, P. Tang, Y. Yang and J. Z. Y. Chen, Soft Matter \textbf{10}, 4674 (2014).

\bibitem{arnold}
A. Arnold, B. Bozorgui, D. Frenkel, B.-Y. Ha and S. Jun, J. Chem. Phys. \textbf{127}, 164903 (2007).

\bibitem{jung}
Y. Jung, S. Jun, and B.-Y. Ha, Phys. Rev. E \textbf{79}, 061912 (2009).

\bibitem{soft2013}
J. Kim, C. Jeon, H. Jeong, Y. Jung and B.-Y. Ha, Soft Matter \textbf{9}, 6142 (2013).


\bibitem{dekker}
D. J. Bonthuis, C. Meyer, D. Stein, and C. Dekker, Phys. Rev. Lett. \textbf{101}, 108303 (2008).

\bibitem{binder2008}
D. I. Dimitrov, A. Milchev, K. Binder, L. I. Klushin, A. M. Skvortsov, J. Chem. Phys. \textbf{128}, 234902 (2008).

\bibitem{slit-revisit}
J. Tang, S. L. Levy, D. W. Trahan, J. J. Jones, H. G. Craighead, and P. S. Doyle, Macromolecules \textbf{43}, 7368 (2010).

\bibitem{semiflexible}
Y.-L. Chen, Y.-H. Lin, J.-F. Chang, and P.-K. Lin, Macromolecules \textbf{47}, 1199 (2014).

\bibitem{extended}
L. Dai, J. van der Maarel, and P. S. Doyle, Macromolecules \textbf{47}, 2445 (2014).

\bibitem{ssDNA}
F. Latinwo and C. M. Schroeder, Soft Matter \textbf{7}, 7907 (2011).

\bibitem{joanny}
J. -L. Barrat and J. -F. Joanny, "Theory of polyelectrolyte solutions", in Advances in Chemical Physics, Polymeric Systems, p. 1 (1997).

\bibitem{tension}
O. A. Saleh, D. B. McIntosh, P. Pincus, and N. Ribeck, Phys. Rev. Lett. \textbf{102}, 068301 (2009);
C. Hyeon, G. Morrison, D. L. Pincus, and D. Thirumalai, Proc. Natl. Acad. Sci. USA \textbf{106}, 20288 (2009).

\bibitem{pre2013} 
L. I. Klushin \emph{et al}, Phys. Rev. E \textbf{87}, 022604 (2013).

\bibitem{nikoofard}
N. Nikoofard and H. Fazli,  Phys. Rev. E \textbf{83}, 050801(R) (2011).

\bibitem{rubinstein}
M. Rubinstein, and R. H. Colby, \emph{Polymer Physics}, Oxford University Press, Oxford (2003).

\bibitem{pre2010}
H. Uemura, M. Ichikawa, and Y. Kimura, Phys. Rev. E \textbf{81}, 051801 (2010).

\bibitem{severe}
H. Liu \emph{et al}, Science \textbf{327}, 64 (2010);
T. Matsuoka \emph{et al}, Nano Lett. \textbf{12}, 6480 (2012);
Y. Kim \emph{et al}, Lab Chip \textbf{11}, 1721 (2011).

\bibitem{spring}
J. Tang \emph{et al}, Macromolecules \textbf{43}, 7368 (2010).

\bibitem{espresso}
H. J. Limbach, A. Arnold, B. A. Mann, and C. Holm, Comput. Phys. Commun. \textbf{174}, 704 (2006).

\bibitem{khalilian}
N. Nikoofard, H. Khalilian, and H. Fazli, J. Chem. Phys. \textbf{139}, 074901 (2013).




\end{thebibliography}
\end{document}